# Evaluating the Business Value of CPOE for Cancer Care in Australia: A Resource Based View Perspective

**Peter Haddad**
Epworth HealthCare and Deakin University
Melbourne, Australia
Email: peter.e.haddad@gmail.com

**Jonathan L. Schaffer**
Cleveland Clinic
Ohio, USA
Email: schaffj@ccf.org

**Nilmini Wickramasinghe**
Epworth HealthCare and Deakin University
Melbourne, Australia
Email: nilmini.work@gmail.com

## Abstract

Today, cancer is one of the leading causes of death throughout the world. This threatening disease has huge negative impacts, not only on quality of life, but also on the healthcare industry, whose resources are already scarce. Thus, finding new approaches for cancer care has been a central point of interest during the last few decades. One of these approaches is the use of computerised physician order entry (CPOE) systems, which have the potential to provide more effective and efficient patient centric cancer care. This paper serves to examine the business value of an American CPOE in an Australian context. This is achieved by using our specifically designed tool to evaluate the business value of IT in the healthcare in combination with a resource based view perspective. Our results show that the system has a number of enabling resources to generate business value subject to having other resources.

**Keywords:** Business Value of IT, Computerised Physician Order Entry (CPOE), Information Systems (IS), Oncology, Resource-based View

## 1. Introduction

Similar to the global trend, cancer is one of the leading causes of death in Australia. According to the Australian Institute of Health and Welfare figures, more than 43,000 Australian people died in 2012 from cancer, which counts about 30% of total deaths in Australia (AIHW 2012). This high number of deaths has had direct and indirect impacts, not only on the Australian society, but also on the economy; cancer treatments cost more than $4.5 billion in direct health system costs, which represents about 6.9% of the total expenditure on the healthcare system in Australia (AIHW 2013).

The safety of medication is highly significant when using anti-cancer therapy. At the same time, cancer treatment is complex and leaves limited margins for errors. In their analysis for errors may happen during cancer treatments, COSA (2008) note that while overdosage can results in death, underdosage can have adverse sequences that affect both the patient outcomes and the management of the disease.

In their explanation of reasons behind errors that may happen during the cancer therapy, COSA (2008) cited a number of reasons that can be classified under three categories: procedural, technical, and behavioural.

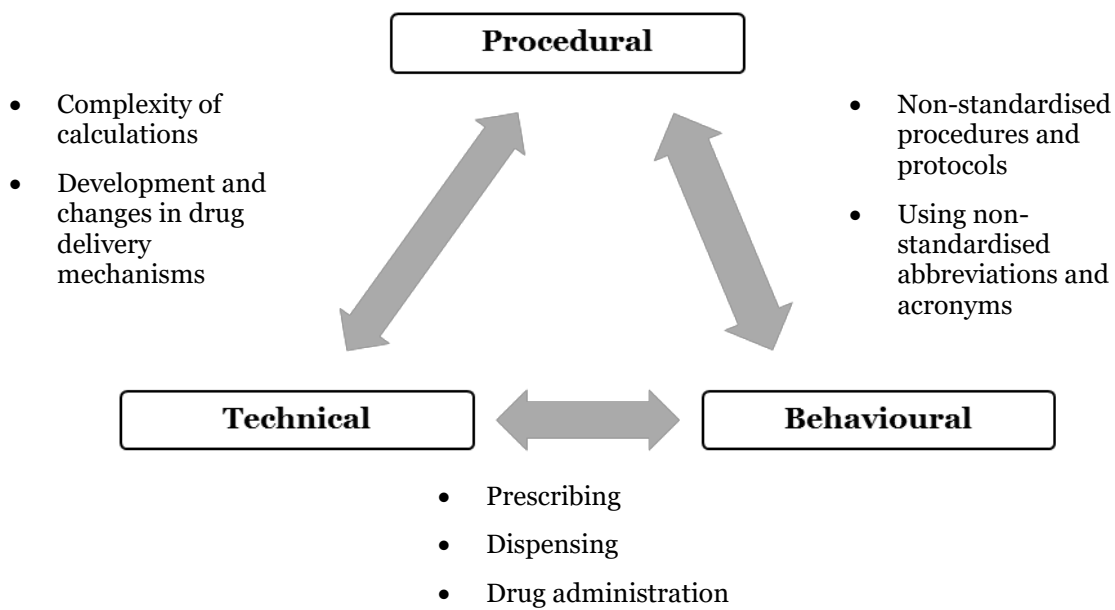

*Figure 1: Main reasons for errors during cancer therapy (adopted from COSA (2008))*

The use of information systems (IS) for oncology purposes has been shown as a key to enhance cancer care (Gandhi et al. 2013). More specifically, the use of computerised physician order entry (CPOE) has widely been agreed upon as an important tool to improve oncology outcomes (Jacobson et al. 2009).

This study serves to examine the use of an American CPOE for cancer care in the Australian context, and its role in generating business value. The remainder of this paper is structured as follows. The next section provides a brief introduction to the concept of business value of IT and the Business Value of IT in Healthcare Model, then a brief summary of the resource-based view theory and its use in this research, followed by the research methodology, results, discussion, and conclusion.

## 2. The Business Value of IT in Healthcare: A Conceptual Model

Nowadays, we are witnessing a proliferation of IS that claim to have positive impacts on healthcare quality, patient outcomes, and financial performance (Haddad and Wickramasinghe 2014). The increased use of IS in the healthcare have called for in-depth examinations of the impact of these IS on the overall performance of the healthcare (Haddad et al. 2014a).

In their review of the business value of IT in the service industry, Melville et al. (2004) used this term to refer to the impacts of IT on the organizational performance, including cost reduction, profitability improvement, productivity enhancement, competitive advantage, inventory reduction, and other measures of performance. Nevertheless, the literature still lacks in-depth analyses of the business value of IT in the healthcare industry (Haddad and Wickramasinghe 2015).

It is important to emphasize that business value of IT is not a value by itself; rather, it is a model that suggests how value might be generated by implementing different IT solutions (Haddad et al. 2014b).

To address this issue in the healthcare industry, we designed the Business Value of IT in Healthcare Model (Figure2) as a tool to examine the business value of different IS in the healthcare industry. It is based on the IT Portfolio theory (Weill and Broadbent 1998) and the Healthcare Delivery Enterprise (Rouse and Cortese 2010).

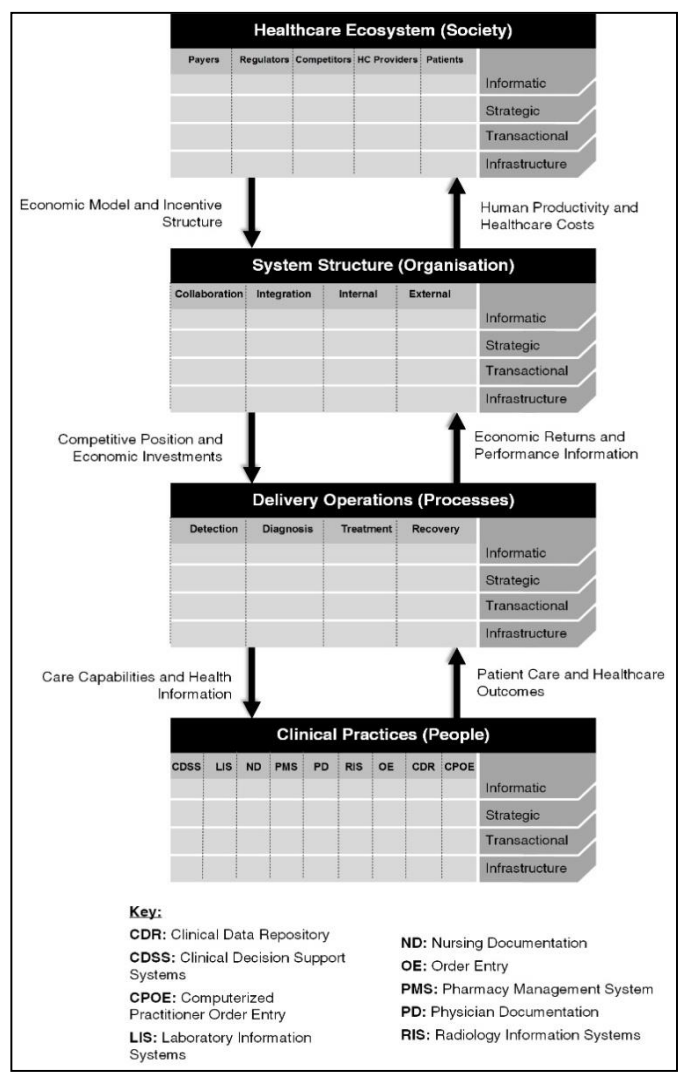

*Figure 2: The Business Value of IT in Healthcare Model*

This tool has proven to be roust and flexible to cover a wide range of IS. A summary of the use of this tool can be found elsewhere (Haddad et al. 2015). For the purpose of this paper, we use this model to examine the business value of a CPOE as aforementioned.

## 3. The Resource-Based View Theory

One of the main arguments of the resource-based theory is that firms may possess resources that can be categorised in two subsets: the first enables to achieve competitive advantages, while the second leads to superior long-term performance (Wernerfelt 1984; Barney 1991; Grant 1991; Wade and Hulland 2004).

As a precise understanding of this theory depends on a precise definition of "resources", Wade and Hulland (2004) defined resources as 'assets and capabilities that are available and useful in detecting and responding to market opportunities or threats' (p.109). While assets can serve as inputs to a process, or as the outputs of a process (Srivastava et al. 1998; Teece at al. 1997), capabilities can transform inputs into outputs of greater worth (Wade and Hulland 2004). Table 1 summarises the different aspects of assets and capabilities from the resource-based view point of view.

| Resources | Description | Reference |
|---|---|---|
| **Assets** | Tangible:<br>IS hardware<br>Network infrastructure | Wade and Hulland 2004 |
|  | Intangible<br>Software patents<br>Strong relationships with vendors | Hall 1997;<br>Srivastava et al. 1998 |
| **Capabilities** | Skills<br>Managerial<br>Technical<br>Processes<br>Systems development<br>Systems integration | Wade and Hulland 2004 |

*Table 1: The 'assets' and 'capabilities' from the resource-based view perspective*

The use of this theory to back this study can be justified from its ability to 1) provide a 'cogent framework to evaluate the strategic value of information systems resources (Hall 1997), 2) its nature in terms of classifying IS into groups based on their separate influence on the organisational performance, which crosses with the main theory used to build our conceptual model i.e. IT Portfolio (Weill and Broadbent 1998).

Thus, by using the lenses provided by the theory of resource-based view, we expect to better understand the strategic value of the studied CPOE, and which assets and/or capabilities would help or prevent generate business value from this system.

## 4. Research Methodology

This paper represents a part of a bigger project to evaluate the business value of IT in the healthcare industry. In this sense, and for the purposes of this paper, we aim at examining the business value of the studied CPOE in cancer care as a clinical practice in the Australian healthcare context.

The aim of this research is to explore the possibilities of the two well formulated theories, i.e. IT Portfolio and Resource-based View theories, to examine different IS/IT in different contexts. Due to this exploratory nature of this research, qualitative approach deemed appropriate (Myers 2009).

To operationalise this research, a case study is used. This has proven to be prudent in similar studies, where deeper understanding of a phenomenon within the selected organisation is both needed and achievable through the insights of key informants within the studies organisation (Yin 2003; Fletcher and Plakoyiannaki 2011).

### 4.1 Data Collection

To operationalise this study, we use a qualitative research approach. The data were collected using two sources: 1) the first is a series of 28 semi-structured interviews with clinicians, executives and IT professional at an Australian case study; 2) the data collected for the study "Identifying Key Success Factors for the Adoption and Implementation of computerised physician order entry systems into the Australian Private Healthcare Sector' (Wickramasinghe et al. 2015), in which an online survey targeted 25 oncologists, 9 executive and non-users that have an access to the CPOE system, and 53 users of the CPOE system (non-oncologists).

The collected data from both sources were adapted according to the terminologies used in the two used theories in this research. i.e. IT portfolio and resource-based theories. The result of this process is presented in Table 2.

| COPE Resources/elements | IT Portfolio | Resource-based View |
|---|---|---|
| Hardware | Infrastructural IT | Tangible assets |
| Network hardware and configuration | Infrastructural IT | Tangible assets |
| IT Human resources | Infrastructural IT | Capabilities (technical) |
| Automating prescribing | Transactional IT | Intangible assets |
| Electronic dispensing | Transactional IT | Capabilities (technical) |
| Creating information-based knowledge | Informational IT | Capabilities (technical/managerial) |
| Managing relationships with vendors | Infrastructural IT (Human resources) | Intangible assets |

*Table 2: Mapping the studied CPOE to IT portfolio and resource-based theories*

### 4.2 The Case study

The case study is XYZ Hospital, which is Victoria's largest not-for-profit private health care group, renowned for excellence in diagnosis, treatment, care and rehabilitation. Through a number of locations across the Melbourne metropolitan area, XYZ is an innovator in Australia's health system, embracing the latest in evidence-based medicine to pioneer treatments and services for patients. In 2013-2014, XYZ had 132,969 patient admissions, 85,207 operations and more than 26,500 emergency department attendances. Cancer treatment at XYZ is one of the major areas of interest and "business", with about 20% of the overall number of patients are receiving cancer treatment/ therapy. This, combined with the strategic planning at XYZ to be ahead in cancer treatment in the Asia pacific region, have created the foundation of investing in a comprehensive oncology management information system, which is described in the following section.

### 4.3 The CPOE

The examined CPOE system in this research is an American system that offers comprehensive information and image management system for the purposes of oncology treatment. It originally offers this services through three highly specialised and sophisticated modules: radiation, medical, and surgical modules. While the first module was purchased and implemented at our case study about two years ago, the second has just been purchased, while, the last has not been acquired yet.

The aim of implementing this system is to replace the manual patients' scheduling by electronic means, which would help make the processes of admission, treatment, and post treatment care plans smoother and with less errors or delay.

## 5. Results

The following section presents some basic statistics about the three groups of participants in the online survey, while the next section is dedicated to present their insights about the studied CPOE. All participants stated they were comfortable to very comfortable in using computers to perform their daily tasks.

### 5.1 Demographic and Professional Results

As mentioned before, the online survey targeted three groups of professionals: oncologists, executives and non-users, and the daily users of the system.

#### 5.1.1 The Oncologists Group

In the group of oncologists, we had 89% of the participants males, and 11% females (Figure 3a), and the majority of them within age group 30-39 years (Figure 3b). Their profession is apparently medical team (100%).

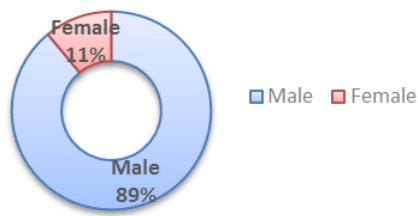
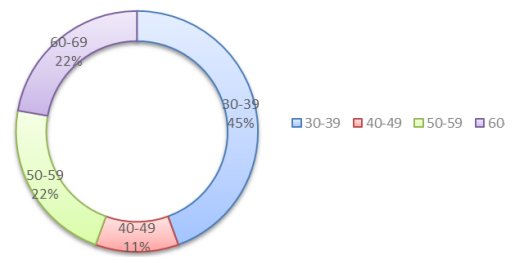

*Figure 3a: Gender within the oncologists group*   *Figure 3b: Age within the oncologists group*

#### 5.1.2 The Executives and Non-users Group

For the executives and non-users group, we had 60% of the participants females and 40% males (Figure 3c), while the age distribution was 60% of them in the age group 40-49 and 40% between 50-59 years (Figure 3d). In terms of their professions, 60% of the executives and non-users came from administration roles, 20% of them came from medical roles, and the remaining 20% came from other roles, primarily IT and finance (Figure 3e).

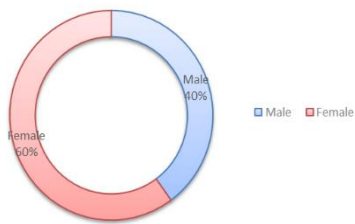 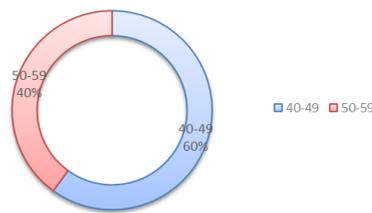 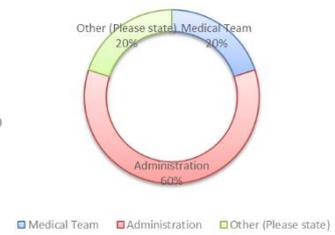

*Figure 3c: Gender within the executives group*　　*Figure 3d: Age within the executives group*　　*Figure 3e: Profession within the executives group*

### 5.1.3 The Users Group

For the users group, We had a similar gender distribution to executives group, i.e. 60% females and 40% males, while we had a different pattern for age distribution (Figure 3f), and profession distribution (Figure 3g). 50% of the 'other' were IT, and 50% finance.

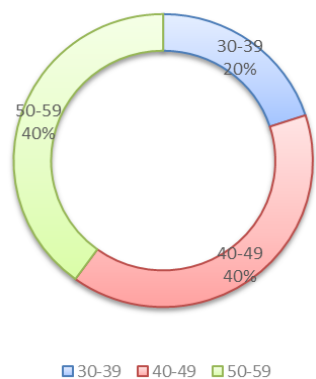 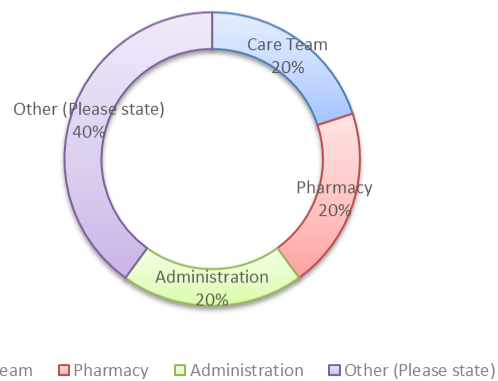

*Figure 3f: Age within the users group*　　*Figure 3g: Profession within the users group*

### 5.2 The Business Value of the CPOE

To evaluate the business value of the CPOE system, we first attempted to understand its main components and how this maps to the Business Value of IT in Healthcare Model. Most of the findings come from the semi-structured interviews with clinicians, executives, and IT personnel at XYZ Hospital. The result of this mapping is presented in Table 3.

| Perspective | Components | CPOE |
|---|---|---|
| IT Portfolio | Infrastructure | Internet, Intranet, Computers, Servers, Databases |
|  | Transactional | Data entry/ input, like identification, progress notes, medication scheduling, discharge checklist treatment plans |

|  |  |  |  |
|---|---|---|---|
|  | Informational |  | Producing and sharing information on treatment plans and medication scheduling |
|  | Strategic | ✓ | Smart medication scheduling |
| Healthcare Delivery | Healthcare Ecosystem |  | The system works in the Australian healthcare ecosystem, and it targets both public and private hospitals. In this study we examine it in the private sector. |
|  | System Structure |  | The system seems to require reengineering healthcare processes to help generate business value. This includes both internal and inter-organisational processes. |
|  | Delivery Operations | ✓ | Detection |
|  |  | ✓ | Diagnosis |
|  |  | ✓ | Treatment |
|  |  | ✓ | Recovery |
|  | Clinical Processes | ✓ | CPOE |

*Table 3: Mapping the CPOE system to the BVIT Model*

Across the three groups of the targeted respondents, there were wide agreement that the system has the ability to help prevent medication errors, overview current treatment plans, support decision making and improve the quality of care outcomes, as figure 4 depicts.

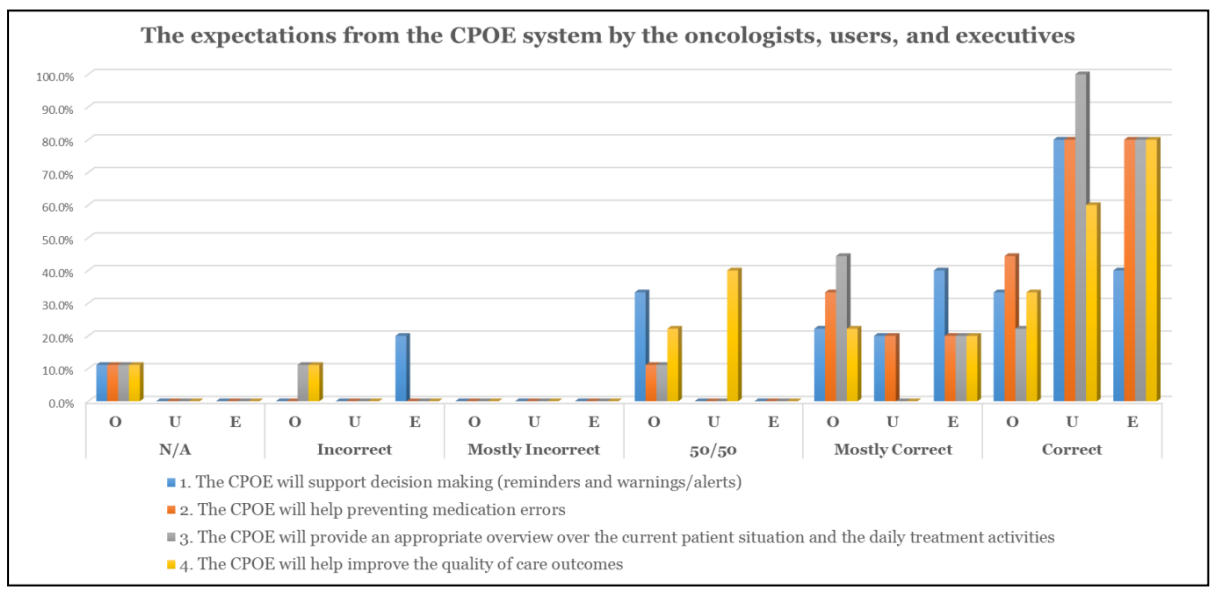

*Figure 4: The expectations from the studied CPOE*

### 5.2.1 Available Resources to Attain Business Value from the CPOE

Analysing the survey and the semi-structured interviews showed that there exists a number of resources that may help generate business value from the CPOE. This includes:

1) A strong and robust infrastructure (networks included) as the IT Business Solutions Manager describes: *"It's strong, it's robust. The guys that we have within the* *infrastructure team are amazing, absolutely amazing. When they build something, they build it to make it resilient"*. This is secured through a team of talented staff with a departmentalization of skills, both managerial and technical as the CIO says: *"The team has got a broad range of skills. I think there's a core which is your infrastructure, your technical services and your service delivery teams, which I have all three, who are caught in the day-to-day mechanics of pieces, hardware, and infrastructure."*

2) Although the system is closed-source software. i.e. the hospital has no access to changing the system without the need for the vendor, this is not a big problem as the hospital and the vendor have strong relationship in the area of oncology treatment, and this has proven to be very helpful for the implementation of the system.

3) Broad range of technical and managerial skills: this can be seen directly from the results coming from the online survey: almost all of the respondents have more than 15 years of experience in the healthcare industry, and feel comfortable/very comfortable to use information systems in fulfilling their day-to-day business.

4) The system fits within the hospital strategy to meet the increasing needs for better oncology treatment. Thus, it gives the hospital both competitive advantages as the XYZ hospital is one of the first healthcare providers Australian-wide to implement this system, and, at the same time, will give the hospital superior long-term performance due to the fact that more than 20% of the patients at XYZ Hospitals are receiving cancer care.

5) The system itself has a number of unique and strong features that enable it to generate business value. According to the survey results this includes: ease of use, its role in increasing the integration between different craft groups (i.e. chemists, oncologists, nurses, etc.), reducing prescribing errors, increasing work efficiency, adhering to guidelines/ protocols, supporting regimen provisions, and improving patient care.

### 5.2.2 Needed Resources to Attain Business Value from the CPOE

Although the studied CPOE system has the aforementioned enabling resources, there are a number of requirements that need to be met in order for this system to better generate sustainable business value:

1) The system fits well within the strategic plan of XYZ Hospital to sustain its pioneer position in oncology care in Australia, but it fits less well within the current information systems used by both the hospital and its visitor medical officers (VMOs), who have invested heavily, both financially and emotionally, in their own information systems. This highlights the need of better process integration and redevelopment measures.

2) This project has represented a 'huge change process' and required clearer policies and procedures that control the interaction with the system, which does not seem to be available right now according to the results from both the survey and the interviews.

3) Although the relationship with the vendor has been described as 'good', 'strong', and 'strategic', there seems to be a real need for better connections with the vendor regarding the required training, better customization of the system with less financial consequences on the hospital. Some aspects of this customisation are: making it accessible online yet secure, adding better reporting capabilities, and better compatibility with other clinical and business IT systems around.

## 6. Discussion

In this study, data analysis shows that the studied CPOE 1) can address the main reasons of errors during cancer therapy, 2) has infrastructural, transactional, and informational components (Weill and Broadbent 1998), 3) has the potential to generate business value for XYZ hospital, 4) has a number of resources that are needed to maximise the business value from it.

The functionality of the system seems to be able to avoid procedural reasons of errors during cancer therapy, as it helps standardize the procedures by better aligning with therapy protocols and standards, and by eliminating the use of non-standardized abbreviations and acronyms, which can also be described as behavioural reasons during the processes of prescribing, dispensing and administration. The system also seems to be very helpful in reducing the complexity of calculations and meeting the dynamically changing nature of drug delivery mechanisms.

In terms of classifying the system against the IT Portfolio (Weill and Broadbent 1998), the data collected through the online survey and the interviews suggests that this system is mainly of informational objectives with both infrastructural and transactional components. Accordingly, this system is expected to play a key role in reducing the cost of cancer therapy, and to achieve quick wins for both the hospital and its patients, as it helps share information between different stakeholders throughout the hospital and make better clinical decisions based on the definition of informational IT (Weill and Broadbent 1998).

The data also reveals that the studied CPOE also has the potential to generate business value for the hospital. i.e. enhance the overall performance of the hospital in the area of oncology. This includes better patient outcomes (safety, experience, quality of care), better work efficiency (doing more with less), and providing cost-effective treatments. These coupled with the "face-value" and the impact of this on the reputation of the hospital as a state-of-the-art system, make investing in this system prudent as our data shows.

Using the lenses of the resource-based view theory shows the usefulness of this theory in evaluating the impact of different IS on the performance of healthcare providers. In this study, we mapped the system and the context in which it is implemented to this theory. This revealed that both assets and capabilities are key to attain business value from the studied CPOE. The hospital seemed to have 1) a robust infrastructure [tangible assets] that supports the functionality of the system [intangible assets], 2) strong relationships with the vendor [intangible assets], 3) a good alignment between the business strategy (pioneer position in cancer care) and the system [capabilities], and 4) a broad range of technical and managerial skills [capabilities]. At the same time, there exists needs for more capabilities by both the hospital and the system. From the system perspective this includes better integration with the current IS throughout the hospital and its VMOs. From the hospital perspective, reengineering the processes to attain better business value from the CPOE system is needed. This includes better change management plans, and even stronger relationships with the vendor, especially for the purpose of better customisation of the system based on the real needs of the hospital.

## 7. Conclusion

This study is set out to evaluate the business value of a CPOE system in the Australian context. A number of strengths, along with rooms of improvement, were identified in this paper by using the Business Value of IT Model and the resource-based view theory. Overall, the system has the potential to generate business value based on its current resources (both assets and capabilities), but this is subject to other required resources.

## 8. References

Australian Institute of Health and Welfare. 2013. "Health system expenditure on cancer and other neoplasms in Australia 2008-2009". Cancer series no. 81. Cat. No. 78. Canberra: AIHW.

Australian Institute of Health and Welfare. 2012. "Cancer incidence projections: Australia, 2011 to 2020". Cancer Series no. 66. Cat. No. CAN 62. Canberra: AIHW.

Barney, J. 2001. "Is the Resource-based "View" a Useful Perspective for Strategic Management Research? Yes," *Academy of Management Review* (26:1), pp. 41-56.

Clinical Oncological Society of Australia (COSA). 2008. "Guidelines for the Safe Prescribing, Dispensing and Administration of Cancer Chemotherapy".

https://www.cosa.org.au/media/1093/cosa_guidelines_safeprescribingchemo2008.pdf. Retrieved 27 July, 2015.

Fletcher, M & Plakoyiannaki, E. 2011. "Case selection in international business: key issues and common misconceptions", in R Piekkari & C Welch (eds.), *Rethinking the Case Study in International Business and Management Research*, Edward Elgar Publishing, Cheltenham, UK, pp. 171-91.

Gandhi, S., Tyono, I., Pasetka, M., & Trudeau, M., 2014. Evaluating an Oncology Systemic Therapy Computerized Physician Order Entry System Using International Guidelines". *Journal of Oncology Practice,* 10(2), E14-E25.

Grant, R. M. 1991. "The Resource-Based Theory of Competitive Advantage: Implications for Strategy Formulation". *California Management Review* (33:1), pp. 114-135.

Haddad, P., Schaffer, J. L., and Wickramasinghe, N. 2015. *Evaluating Business Value of IT in Healthcare: Three Clinical Practices from Australia and the US*. In MEDINFO 2015 August 19-23 Sao Paulo, Brazil: E-Health-enabled Health: Proceedings of the 15th World Congress on Health and Biomedical Informatics (Vol. 216, p. 183). IOS Press.

Haddad, P., & Wickramasinghe, N. 2015. The Use of a Nursing Informatics System as an Exemplar to Investigate Business Value of IT in Healthcare. *Health and Technology*, 5(1), pp.25-33, ISSN: 2190-7188.

Haddad, P., and Wickramasinghe, N. 2014. *Conceptualizing the Business Value of IT in Healthcare to Design Sustainable E-Health Solutions*. AMCIS Savannah August 2014.

Haddad, P., Gregory, M. and Wickramasinghe, N. 2014a. *Evaluating the Business Value of IT in Healthcare in Australia: The Case of an Intelligent Operational Planning Support Tool Solution*. Bled eConference, Bled, Slovenia.

Haddad, P., Gregory, M. and Wickramasinghe, N. 2014b. Business Value of IT in Health Care. In Wickramasinghe et al. (Eds.), *Lean Thinking for Healthcare* (pp.55-82), New York, Springer.

Hall, R. 1997. *Complex Systems, Complex Learning, and Competence Building"*, Wiley, New York, 1997.

Jacobson, J., Polovich M., McNiff K. 2009. American Society of Clinical Oncology/Oncology Nursing Society chemotherapy administration safety standards. *Journal of Clinical Oncology.* 27:5469–5475.

Melville, N., Kraemer, K., & Gurbaxani, V. 2004. Review: Information Technology and Organizational Performance: An Integrative Model of IT Business Value. *MIS Quarterly*, 28(2), 283-322.

Rouse, W. B., & Cortese, D. A. 2010. *Engineering the system of healthcare delivery*. Amsterdam: Amsterdam: IOS Press.

Srivastava, R., Shervani, T., & Fahey, L. 1998. "Market-Based Assets and Shareholder Value: A Framework for Analysis". *Journal of Marketing*, 62(1), 2-18.

Teece, D., Pisano, G., & Shuen, A. 1997. "Dynamic capabilities and strategic management". *Strategic Management Journal*, 18(7), 509-533.

Weill, P., & Broadbent, M. 1998. *Leveraging the new infrastructure: how market leaders capitalize on information technology*. Boston, Mass.: Harvard Business School Press.

Wernerfelt, B. 1984. "A Resource-Based View of the Firm", *Strategic Management Journal* (5), 1984, pp. 171-180.

Wade, M., & Hulland, J. 2004. "Review: the resource-based view and information systems research: review, extension, and suggestions for future research", *MIS Quarterly*, v.28 n.1, p.107-142.

Wickramasinghe, N., Vaughan, S., Haddad, P., Han-Lin, C., 2015. *Identifying Key Success Factors for the Adoption and Implementation of a Chemotherapy Ordering System: A Case Study from the Australian Private Healthcare Sector*. AMCIS 2015 Puerto Rico. August 13-15, 2015 Proceedings. ISBN: 978-0-9966831-0-4.

Yin. 2003. "*Case Study Research: Design and Methods*", 3rd edition, Sage Publications, Newbury Park.

## Acknowledgements:

This project was funded by Varian Medical Systems. All appropriate ethics clearances were obtained.

## Copyright